\documentstyle[preprint,aps,epsf,rotate]{revtex}

\begin{document}
\draft


\title{\LARGE Coulomb Gap and Correlated Vortex Pinning in Superconductors}

\author{Uwe C. T\"auber, $^1$ Hongjie Dai, $^2$
        David R. Nelson, $^1$ and Charles M. Lieber $^2$}

\address{$^1$ Lyman Laboratory of Physics, Harvard University, Cambridge,
              Massachusetts 02138}

\address{$^2$ Division of Applied Sciences and Department of Chemistry,
              Harvard University, Cambridge, Massachusetts 02138}

\date{\today}
\maketitle


\begin{abstract}

The positions of columnar pins and magnetic flux lines determined from a
decoration experiment on BSCCO were used to calculate the single--particle
density of states at low temperatures in the Bose glass phase. A wide Coulomb
gap is found, with gap exponent $s \approx 1.2$, as a result of the long--range
interaction between the vortices. As a consequence, the variable--range hopping
transport of flux lines is considerably reduced with respect to the
non--interacting case, the effective Mott exponent being enhanced from
$p_0 = 1/3$ to $p_{\rm eff} \approx 0.5$ for this specific experiment.

\end{abstract}




The remarkably rich phase diagram of magnetic flux lines in high--temperature
superconductors, especially when subject to point and/or extended pinning
centers, has attracted considerable experimental and theoretical interest
\cite{review}. Understanding the interaction of vortices with defects is
especially important since flux lines must be pinned to mimimize dissipative
losses from flux creep. A promising pinning strategy involves the creation of
linear damage tracks in materials by heavy ion irradiation. These columnar
defects have been found to increase the critical current, as well as to shift
the irreversibility line significantly upwards \cite{pinexp}.

A theory of flux pinning by correlated disorder has been developed to explain
these results, exploiting a formal mapping of the statistical mechanics of
directed lines onto the quantum mechanics of two--dimensional bosons
\cite{nelvin}. In this study, the intervortex repulsion, whose range is
determined by the London penetration depth $\lambda$, was only treated using
approximate, order of magnitude estimates. However, if $\lambda \gg a_0$, where
$a_0 = (4/3)^{1/4} (\phi_0 / B)^{1/2}$ is the average distance between vortices
($\phi_0 = h c / 2 e$ is the elementary flux quantum), the interactions become
effectively long--range, and may lead to important correlation effects.

Indeed, the analogy of flux lines at low temperatures pinned to columnar
defects (Bose glass phase) with two--dimensional localized carriers in doped
semiconductors (Coulomb glass) \cite{cgapth}, suggests that a ``Coulomb'' gap
should emerge in the single--particle density of states (distribution of vortex
pinning energies). Because such a gap will affect significantly the
current--voltage characteristics in a variable range hopping approach
\cite{nelvin}, it is important to estimate the size of the Coulomb gap in the
Bose glass phase and to understand the correlation effects induced by the
intervortex repulsion. The recent successes, moreover, in simultaneously
measuring both the columnar defect and flux line positions \cite{decexp,expdat}
allow for detailed comparison of the spatial correlations found in experiment
and theory. Such a comparison is not feasible in the semiconductor case.


The density of states may be obtained by using a variant of the Monte Carlo
algorithm described by Shklovskii and Efros \cite{cgapth,leesim}. Using the
experimentally determined columnar defect and flux line positions shown in
Fig.~1, we can predict the density of states and the transport characteristics
for this specific sample in the variable range hopping regime, at temperatures
slightly below the depinning transition. We find that the ensuing Coulomb gap
is remarkably large, even in the case $\lambda \approx a_0$, i.e. when the
interactions are relatively short--range. Vortex interactions raise the
effective Mott exponent from the non--interacting result $p_0 = 1/3$ to a value
$p_{\rm eff} \approx 0.5$ for this specific sample. These results imply that
correlation effects strongly enhance the pinning of flux lines to columnar
defects.


We consider the following model free energy for $N$ flux lines, defined by
their trajectories ${\bf r}_i(z)$ as they traverse the sample of thickness $L$,
with the magnetic field ${\bf B}$ aligned along the $z$ axis (perpendicular to
the ${\rm CuO}$ planes), \cite{nelvin}
\begin{eqnarray}
  F = \int_0^L dz \sum_{i=1}^N &&\Biggl\{
    {{\tilde \epsilon}_1 \over 2} \left( {d {\bf r}_i(z) \over dz} \right)^2 +
    {1 \over 2} \sum_{j \not= i}^N V[r_{ij}(z)] \nonumber \\
    &&\quad + \sum_{k=1}^{N_D} V_D[{\bf r}_i(z) - {\bf R}_k] \Biggr\} \quad .
\label{modelf}
\end{eqnarray}
Here $r_{ij}(z) = | {\bf r}_i(z) - {\bf r}_j(z) |$, and $V(r) = 2 \epsilon_0
K_0(r / \lambda)$ is the repulsive interaction potential between the lines; the
modified Bessel function $K_0(x) \propto - \ln x$ for $x \rightarrow 0$, and
$\propto x^{-1/2} e^{-x}$ for $x \rightarrow \infty$. Thus the (in--plane)
London penetration depth $\lambda$ defines the interaction range. The energy
scale is set by $\epsilon_0 = (\phi_0 / 4 \pi \lambda)^2$, and ${\tilde
\epsilon}_1$ is the tilt modulus. Finally, the pinning energy is a sum of $N_D$
$z$--independent potential wells $V_D$ with average spacing $d$ centered on the
$\{ {\bf R}_k \}$, whose diameters are typically $c_0 \approx 100 \AA$, with a
variation induced by the dispersion of the ion beam of $\delta c_k / c_0
\approx 15 \%$. This induces some probability distribution $P$ of the pinning
energies $U_k$, which may be determined from the (interpolation) formula $U_k
\approx (\epsilon_0 / 2) \ln [1 + (c_k / \sqrt{2} \xi)^2]$, where $\xi$ is the
coherence length \cite{nelvin}. E.g., for $\lambda \approx 4200 \AA$ one has
$U_0 \approx 0.67 \epsilon_0$, and $w = \sqrt{\langle \delta U_k^2 \rangle}
\approx 0.1 \epsilon_0$.

As is explained in detail in Ref.~\cite{nelvin}, the statistical mechanics
of the model (\ref{modelf}) can be formally mapped onto a two--dimensional
zero--temperature quantum mechanical problem by using a transfer matrix
approach. In this boson analogy, the real temperature $T$ assumes the role of
$\hbar$ in the quantum problem, and the boson electric field and current
density map on the superconducting current $J$ and the true electric field
${\cal E}$, respectively (see Table I in Ref.~\cite{nelvin}). Thus the roles of
conductivity and resistivity become interchanged.

We are interested in the low--temperature properties of flux lines pinned to
columnar defects, with filling fraction $f = N / N_D = (d / a_0)^2 < 1$, in the
Bose glass phase, where all the vortices are assumed (and found) to be bound to
the pinning centers. For $T$ less than a characteristic fluctuation temperature
$T_1$, \cite{nelvin} one arrives at the classical limit of the corresponding
boson problem ($\hbar \rightarrow 0$), and as the flux lines are now well
separated, the Bose statistics become irrelevant. Furthermore, in this limit
thermal wandering is suppressed, and the flux lines will be essentially
straight; hence the tilt energy in Eq.~(\ref{modelf}) can be neglected. In the
boson representation we eventually have to deal with a time--independent
problem defined by the two--dimensional effective Hamiltonian
\begin{equation}
{ H = {1 \over 2} \sum_{i \not= j}^{N_D} n_i n_j V(r_{ij})
                + \sum_{i=1}^{N_D} n_i t_i \quad , }
\label{effham}
\end{equation}
where $i,j = 1,\ldots,N_D$ denote the defect sites, randomly distributed on the
$xy$ plane. $n_i = 0, 1$ is the corresponding site occupation number, and
(originating in the varying pin diameters) the $t_i$ are random site energies,
whose distribution may be chosen to be centered at ${\overline t} = 0$, with
width $w$ [for simplicity, we assume a flat distribution $P(t_i) = 1 / 2w$ for
$| t_i | \leq w$, and $P(t_i) = 0$ else].


The Hamiltonian (\ref{effham}) is precisely of the form studied in the context
of charge carriers localized at random impurities in doped semiconductors
(Coulomb glass problem) \cite{cgapth,leesim,gapsim}. We remark that it is
equivalent to a two--dimensional random--site, random--field antiferromagnetic
Ising model with long--range exchange interactions, and has, at least to our
knowledge, eluded successful analytical approaches going beyond simplifying
mean--field type considerations \cite{cgapth}, and phenomenological scaling
arguments \cite{ftyexp}. Therefore one has to resort to numerical studies using
suitable Monte Carlo algorithms, as described in Refs.~\cite{cgapth,leesim}.

Basically, starting from a given distribution of $N_D$ defect sites, $N < N_D$
of which are occupied, single--particle energies are calculated according to
$\epsilon_i = \partial H / \partial n_i = \sum_{j \not= i}^{N_D} n_j V(r_{ij})
+ t_i$. The initial configuration is then relaxed by moving single
``particles'' to empty places, until precisely the $N$ lowest energy levels are
occupied ($\epsilon_{\rm max}^1 = {\rm max}_{n_i=1} \epsilon_i < \epsilon_{\rm
min}^0 = {\rm min}_{n_i=0} \epsilon_i$). The ensuing intermediate state is then
probed against all possible single--particle hops from filled to empty defect
sites, with associated energy change $\Delta_{i \rightarrow j} = \epsilon_j -
\epsilon_i - V(r_{ij})$. If any $\Delta_{i \rightarrow j} < 0$, the move from
site $i$ to $j$ is performed, and thus the total energy is decreased. (Note
that this procedure constitutes a zero--temperature algorithm.) Afterwards the
``equilibration'' step has to be repeated, for all the $\epsilon_i$ of course
rearrange upon changes in the occupation numbers $n_i$. Finally, a chemical
potential $\mu$ is calculated (approximately) by $\mu = (\epsilon_{\rm min}^0 -
\epsilon_{\rm max}^1) / 2$. [Formally, this corresponds to adding a term $- \mu
\sum_i n_i$ to Eq.~(\ref{effham}); this chemical potential is related to the
external magnetic field $H$ by $\mu = \epsilon_0 \ln (\lambda / \xi) - H \phi_0
/ 4 \pi - \langle U_k \rangle$, where $\langle U_k \rangle$ includes a small
thermal renormalization of pinning energies.]

By repeating this procedure for different initial configurations, one may then
obtain the density of states $g(\epsilon)$ from the site energy statistics. We
shall use a normalization of the density of states such that $\int g(\epsilon)
d \epsilon = 1 / d^2$. To find the correct ground state for each configuration,
one would in principle have to test the configuration against any simultaneous
$n$--particle hops, $n = 2,\ldots,\infty$. However, previous investigations
have shown that terminating at $n=1$ already yields at least qualitatively
reliable estimates for the energy level distributions
\cite{cgapth,leesim,gapsim}. We have performed many such simulations for a
variety of filling fractions $f$ and values of $\lambda / d$. Details of these
investigations and their results shall be reported elsewhere \cite{taunel}. In
this Letter, we shall instead use these techniques to determine the density of
states {\it directly} from experimental data, obtained as follows.


The positions of columnar defects and flux lines in a ${\rm Bi_2Sr_2CaCu_2O_8}$
(BSCCO) crystal were determined simultaneously by a combined chemical
etching/magnetic decoration approach (for details, see Ref.~\cite{expdat}). A
digitized SEM image of the positions of $N = 162$ flux lines and $N_D = 686$
columnar defects (hence $f \approx 0.24$) for a BSCCO sample irradiated with
corresponding matching field $B_\phi = 118 {\rm G}$ ($d \approx 4400 \AA$)
and decoration field $B = 27 {\rm G}$ ($a_0 \approx 9400 \AA$) is shown in
Fig.~1. The critical and irreversibility temperature under these conditions are
$T_c = 87 {\rm K}$ and $T_{\rm irr} = 81 {\rm K}$, respectively. Assuming the
flux line distribution is frozen in at $T_{\rm irr}$, we estimate that the
effective London penetration depth is $\lambda(T_{\rm irr}) \approx 4200 \AA$
\cite{efflam}, and $\lambda / d \approx 0.96$.


In Fig.~2 the structure factor $S(q)$ for the flux lines is depicted, as
obtained from $S({\bf q}) = {1 \over N} \sum_{i,j}^N e^{i {\bf q}({\bf r}_i -
{\bf r}_j)}$ by averaging over directions in Fourier space (thick line).
Clearly the flux line distribution is highly correlated, with $S(q)$ displaying
a peak at $q_0 a_0 \approx 2 \pi$. We have used the experimental defect and
flux line positions as an initial state for the Monte Carlo routine. In order
to minimize boundary effects, we have kept the configuration fixed in a frame
extending $10 \%$ (which amounts to $\approx 3 \lambda$) inwards from each of
the rectangular boundaries. This leaves $464$ sites and $106$ displacable
particles for the simulation. For vanishing site randomness ($w = 0$), it turns
out that about $20 \%$ of the flux lines are moved in the course of the energy
minimization process, which may in part be attributed to finite size effects,
and also to the unaccounted variation in sample pinning energies. Using the
more realistic value $w = 0.1 \epsilon_0$ instead, we find typically $40 \%$
changes with respect to the original experimental distribution. But, as can be
seen in Fig.~2, in both cases the highly correlated character of the flux line
distribution is preserved, although the height of the peak at $q_0$ decreases
upon increasing $w$. We have found, however, that substantially stronger
disorder would destroy the peak in $S(q)$. Similarly, for considerably lower
values of $\lambda / d$ ($\lambda / d \leq 0.2$), the spatial correlations also
disappear, because the random site energies would then dominate over the
interactions. The correlations we do find strongly support the assumption that
the effective London penetration depth is the larger value $\lambda_{\rm eff} =
\lambda(T = T_{\rm irr})$.

The density of states for the $464$ ``inner'' sites, as obtained from averaging
over $100$ runs with different assignments of random site energies, drawn
though from the same distribution $P(t_i)$ with width $w = 0.1 \epsilon_0$, and
using $\lambda_{\rm eff}$ for the interaction range, is shown in Fig.~3.
(Results for the density of states are remarkably insensitive to the precise
value $w$.) The Coulomb gap, separating the occupied and empty energy levels,
is remarkably wide, its width amounting to $\approx 0.5 \epsilon_0$ at half
maximum [$d^2 g(\epsilon) \approx 0.8$], i.e., almost a third of the total
width of $g(\epsilon)$ there. Near its minimum, this pseudogap may be described
by the formula
\begin{equation}
{ g(\epsilon) \propto | \epsilon - \mu |^s \quad , }
\label{gapexp}
\end{equation}
with a gap exponent $s \approx 1.2$ for the specific parameter values here. In
the immediate vicinity of the chemical potential, this power law is smeared
out, and $g(\mu)$ is actually finite (but very small) due to the finite range
of the interactions. We remark that in the limit $\lambda \rightarrow \infty$
and for small filling, $s \approx 3$ may be reached \cite{taunel}.


{}From this single--particle density of states, we may now infer the transport
properties in the variable--range hopping regime by minimizing the free energy
of two superkinks of size $R$ and separation $Z$, $\delta F_{\rm SK} = 2 E_K R
/ d + Z \Delta(R) - f_L R Z$ \cite{nelvin}. Here, the first term consists of
the energy of the double kink, with $E_K = \sqrt{{\tilde \epsilon}_1 U_0}d$,
and the third one derives from the Lorentz force $f_L = \phi_0 J / c$ induced
by the current $J$. The second contribution stems from the fact that for a hop
of distance $R$, the available energy states lie in the interval
$[\mu,\Delta(R)]$, where in $D$ dimensions (here $D=2$) $\Delta(R)$ is
determined by the equation $\int_\mu^\Delta g(\epsilon) d \epsilon = R^{-D}$.
Minimizing first for $J = 0$ gives the longitudinal extent of the kink to be
$Z^* = - 2 E_K / d (\partial \Delta / \partial R)_{R^*}$. To first order in
$J$, one subsequently arrives at $J \phi_0 / c = \Delta(R^*) / R^*$, and thus
$\delta F^* = 2 E_K R^*(J) / d$ is the result for the optimized free energy.
The latter finally enters the resistivity $\rho$ as an energy barrier in an
activation factor, ${\cal E} = \rho_0 J \exp(- \delta F^* / k_B T)$. Using
$g(\epsilon)$ of Fig.~3 then amounts to studying transport slightly below the
depinning temperature, at $T = 77 {\rm K}$, say. Note that for this field range
${\tilde \epsilon}_1 \approx \epsilon_0$ and thermal renormalizations of
pinning energies become relevant only for $T_1 \approx 0.9 T_c$ ($\approx 78
{\rm K}$ here) [3]. Thus, all temperatures $T < T_1$ may be considered as
``low''.

In the regime where Eq.~(\ref{gapexp}) holds, the final result from these
considerations for the highly nonlinear current--voltage characteristics may be
cast in the form
\begin{equation}
{ {\cal E} \approx \rho_0 J
           \exp \left[ - (2 E_K / k_B T) (J_0 / J)^p \right] \; , }
\label{ivchar}
\end{equation}
where $p$ is an exponent generalizing Mott's law [$p_0 = 1 / (D+1)$], which is
valid in the case of vanishing interactions. For long--range interactions
producing a Coulomb gap of the form (\ref{gapexp}), one finds $p = (s + 1) /
(D + s + 1)$. Fig.~4 shows a $\log$--$\log$ plot of the function $R^*(J) / d$
vs. $j = J \phi_0 d / 2 \epsilon_0 c$, derived from the density of states in
Fig.~3, as compared to a similarly calculated curve with the vortex repulsion
being switched off. While in the latter case the result is indeed a straight
line with slope $-1/3$, interactions considerably enhance the pinning by
raising the effective Mott exponent to $p_{\rm eff} \approx 0.5$ for not too
low values of $J$. For $J \rightarrow 0$ the cutoff of the interaction at
$\lambda$ reduces $p_{\rm eff}$ somewhat.


Similar to the gap index $s$, from which it is derived, $p$ really should not
be understood as a universal number, but rather as some effective exponent
$p_{\rm eff}$ conveniently describing the IV characteristics. Its value in
general depends on both the filling $f$ and the interaction range $\lambda /
d$; its maximum value $p_{\rm eff} \approx 0.68$ is reached for $\lambda
\rightarrow \infty$ and small $f$ \cite{taunel}. These results clearly rule out
the mean--field estimate, which would yield $s = D / \sigma - 1$ and $p = 1 /
(1 + \sigma)$ for a potential $V(r) \propto r^{-\sigma}$ ($\sigma < D$; a
logarithmic interaction is recovered in the limit $\sigma \rightarrow 0$).
Rather, they seem consistent with the analysis by Fisher, Tokuyasu, and Young,
\cite{ftyexp} who relate $p$ with the stiffness exponent $\Theta \approx -0.5$
and a fractal index $1 \leq \psi_Q \leq 2$ for the supposedly equivalent gauge
glass model. Their scaling relation $p = 1 / (1 + |\Theta|/\psi_Q)$, indeed
yields $p \approx 2/3$ for $\psi_Q \approx 1$. It is, however, not yet settled
if universality does apply in the purely long--range limit, and even the
results by M\"obius and Richter, \cite{gapsim} who were able to simulate the
$1/r$ Coulomb problem in two and three dimensions for very large systems, have
apparently not reached the fully asymptotic regime. Also, our derivation of
$R^*(J)$ from $g(\epsilon)$ of course constitutes an approximation which
neglects some subtle correlations, e.g. the spatial clustering of those sites
which are energetically close to $\mu$, \cite{leesim,taunel}, and may be
subject to corrections in the limit $J \rightarrow 0$.


In summary, we have demonstrated that the vortex--vortex repulsion can lead to
remarkable correlations both in real space and in the single--particle density
of states, even for $\lambda \approx a_0$, which is easily accessible in
experiment. An important consequence of these correlation effects is the
drastic enhancement of flux line pinning to columnar defects in the Bose glass
phase, whenever $\lambda \geq a_0$.


We benefitted from discussions with A.L.~Efros, D.S.~Fisher, T.~Hwa, P.~Le
Doussal, and V.M.~Vinokur.
This work was supported by the National Science Foundation, primarily
by the MRSEC Program through Grant DMR-9400396, and through Grants
DMR-9417047 and DMR-9306684.
U.C.T. acknowledges support from the Deutsche Forschungsgemeinschaft (DFG)
under Contract Ta.~177/1-1.



\newpage

\begin{figure}
FIG.~1. Positions of columnar pins (open circles) and flux lines (filled
        circles), as obtained from the experiment.
\label{fig1}

FIG.~2. Flux line structure function $S(q)$ as obtained from the experiment
        (thick line), and from the simulation with $w = 0$ (dashed), and
        $w = 0.1 \epsilon_0$ (thin line), averaged over $100$ different
        assignments of random site energies.
\label{fig2}

FIG.~3. Normalized density of states $G(E) = d^2 g(\epsilon)$ as function of
        the single--particle energies $E = \epsilon / 2 \epsilon_0$, averaged
        over $100$ runs ($\lambda / d = 0.96$, $w = 0.1 \epsilon_0$). The mean
        chemical potential is $\mu / 2 \epsilon_0 = 0.65$.
\label{fig3}

FIG.~4. Double--logarithmic plot of the exponential factor $R^*(J) / d$ for
        Mott variable--range hopping vs. $j = J \phi_0 d / 2 \epsilon_0 c$, as
        obtained from the density of states in Fig.3 (filled circles),
	compared to the result of the non--interacting case (open circles).
\label{fig4}
\end{figure}


\begin{references}

\bibitem{review} For a recent review, see G.~Blatter, M.V.~Feigel'man,
                 V.B.~Geshkenbein, A.I.~Larkin, and V.M.~Vinokur,
		 Rev. Mod. Phys. {\bf 66}, 1125 (1994).

\bibitem{pinexp} See, e.g., L.~Civale {\it et al.},
		 Phys. Rev. Lett. {\bf 67}, 648 (1991);
                 M.~Konczykowski {\it et al.},
		 Phys. Rev. B {\bf 44}, 7167 (1991);
                 R.C.~Budhani, M. Suenaga, and S.H. Liou,
		 Phys. Lett. {\bf 69}, 3816 (1992).

\bibitem{nelvin} D.R.~Nelson and V.M.~Vinokur,
		 Phys. Rev. B {\bf 48}, 13060 (1993); and references therein.

\bibitem{cgapth} See, e.g., B.I.~Shklovskii and A.L.~Efros,
                 {\it Electronic Properties of Doped Semiconductors}
                 (Springer, New York, 1984); and references therein.

\bibitem{decexp} S.~Behler {\it et al.},
		 Phys. Rev. Lett. {\bf 72}, 1750 (1994);
		 Z. Phys. B {\bf 94}, 213 (1994).

\bibitem{expdat} H.~Dai, S.~Yoon, J.~Liu, R.C.~Budhani, and C.M.~Lieber,
                 Science {\bf 265}, 1552 (1994); and references therein.

\bibitem{leesim} J.H.~Davies, P.A.~Lee, and T.M.~Rice,
		 Phys. Rev. Lett. {\bf 49}, 758 (1982);
		 Phys. Rev. B {\bf 29}, 4260 (1984);
		 E.I.~Levin, V.L.~Nguen, B.I.~Shklovskii, and A.L.~Efros,
		 Sov. Phys. JETP {\bf 65}, 842 (1987)
		 [Zh. Eksp. Teor. Fiz. {\bf 92}, 1499 (1987)].

\bibitem{gapsim} A.~M\"obius, M.~Richter, and B.~Drittler,
		 Phys. Rev. B {\bf 45}, 11568 (1992).

\bibitem{ftyexp} M.P.A.~Fisher, T.A.~Tokuyasu, and A.P.~Young,
  		 Phys. Rev. Lett. {\bf 66}, 2931 (1991).

\bibitem{taunel} U.C.~T\"auber and D.R.~Nelson, unpublished.

\bibitem{efflam} Upon assusing $\lambda_0 = \lambda(T=0) \approx 2100 \AA$ and
                 the two--fluid formula, $\lambda(T) = \lambda_0 /
                 [ 1 - (T/T_c)^4 ]^{1/2}$, we find that $\lambda(T_{\rm irr})
                 \approx 2 \lambda_0$.

\end{references}
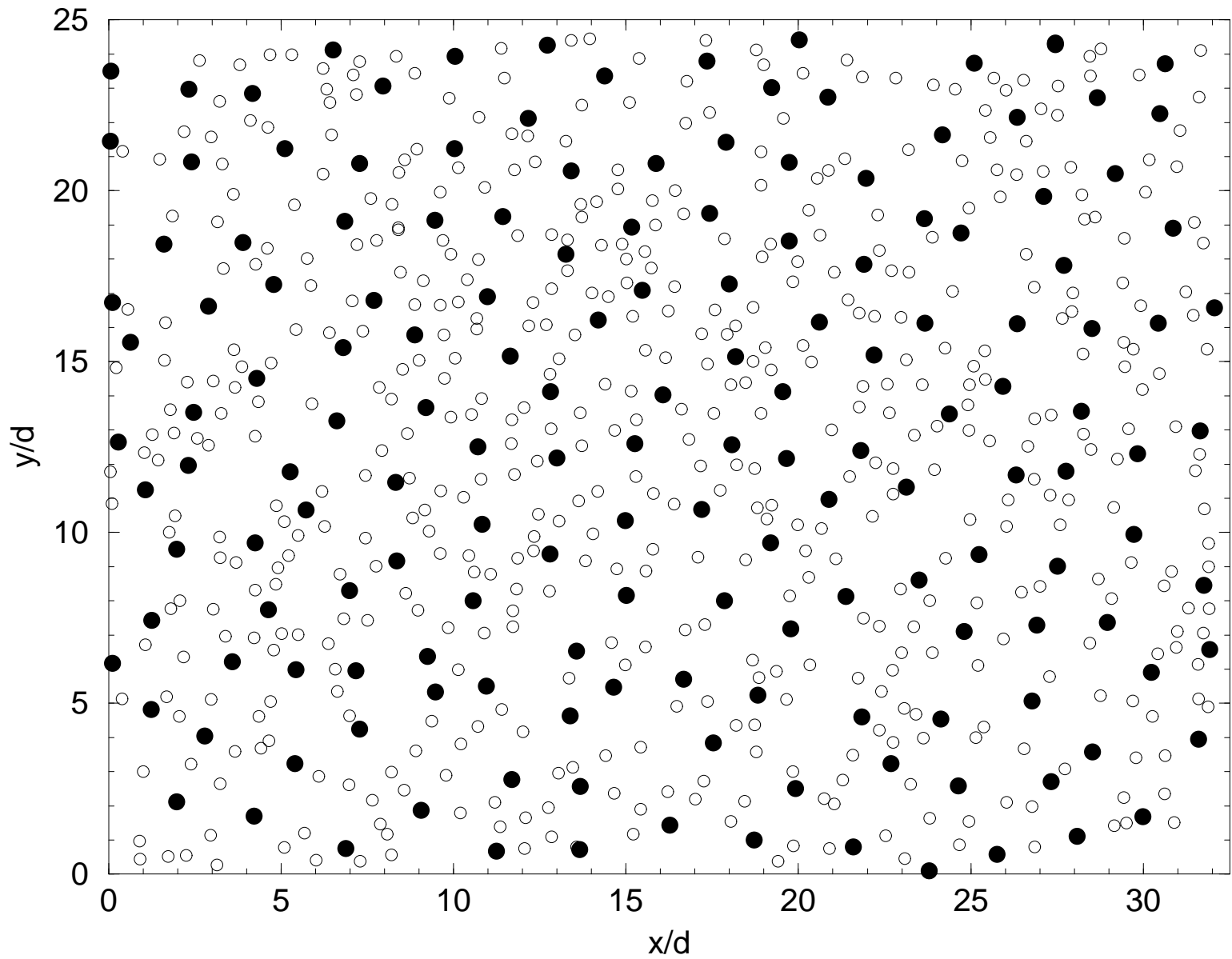
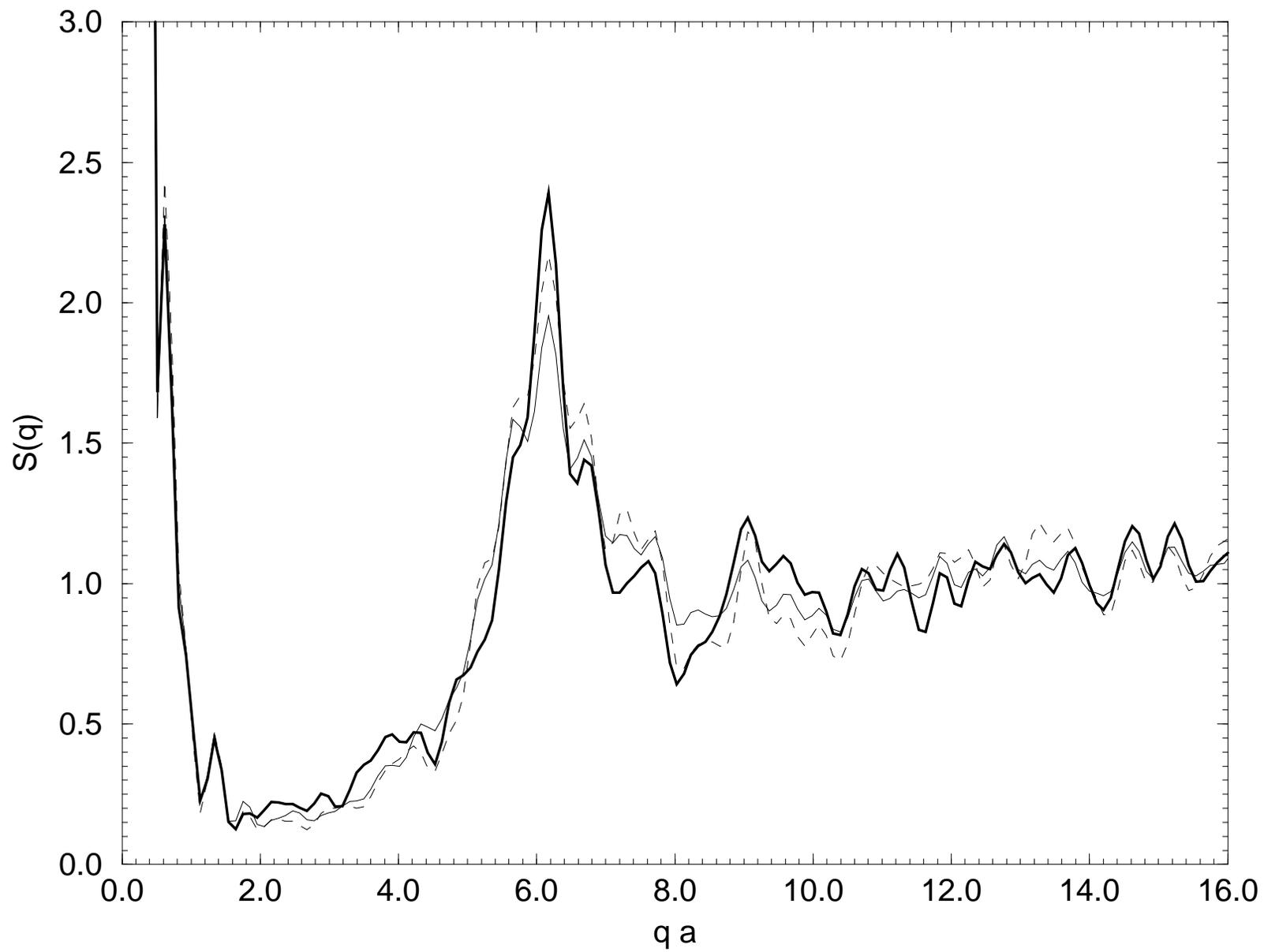
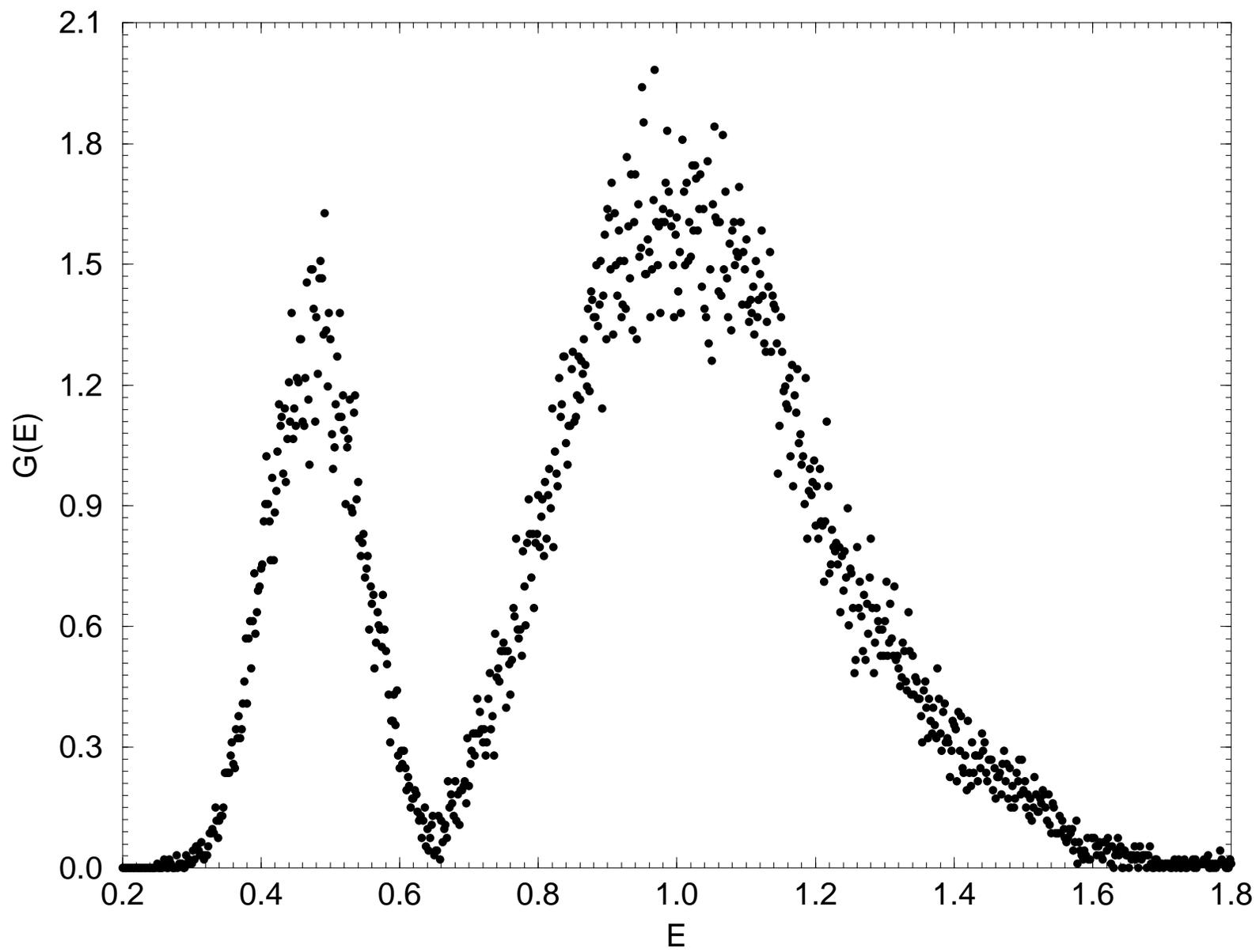
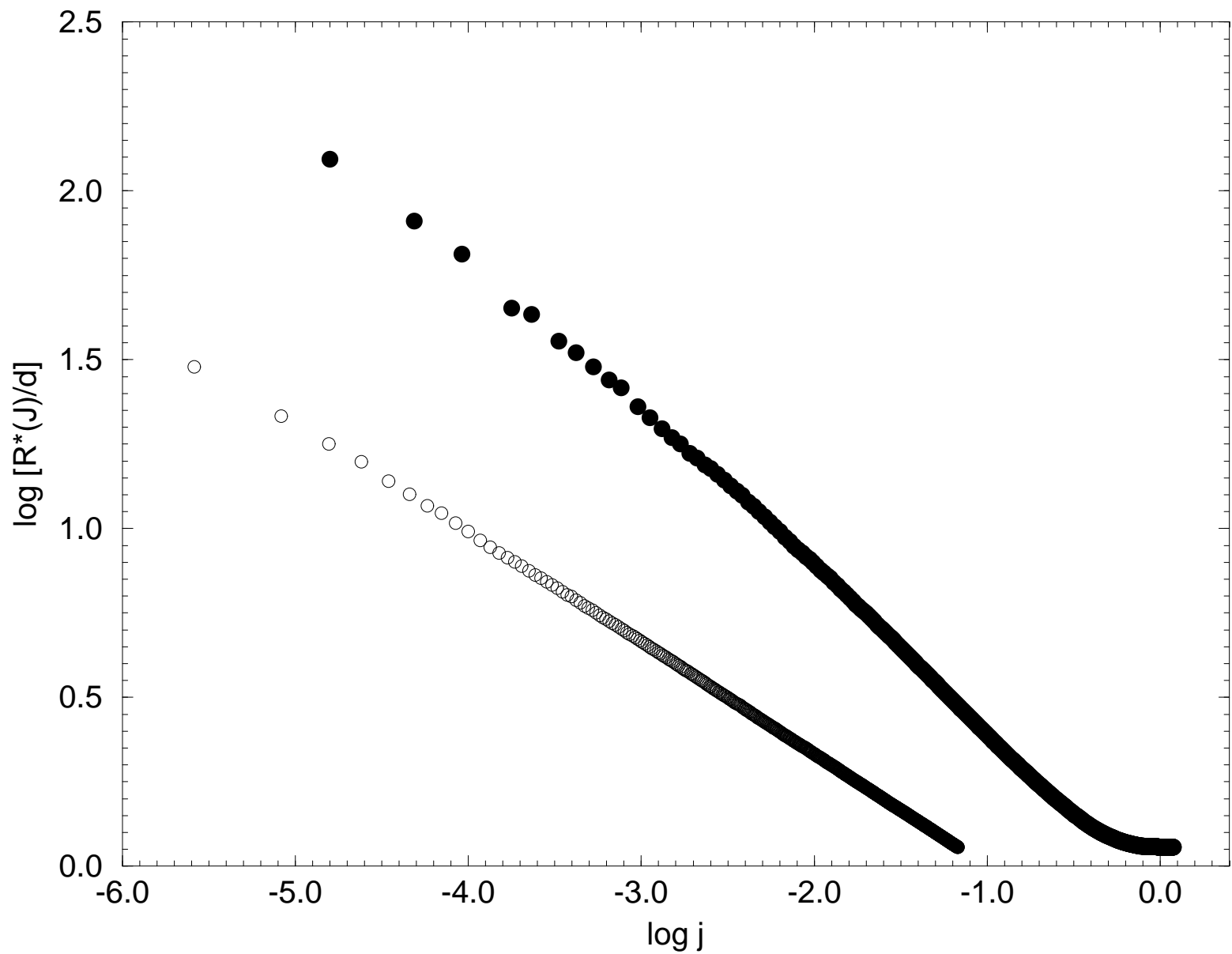
\end{document}